\documentclass[11pt]{article}
\usepackage{chemsym,amssymb}
\begin{document}
\title{Theory of colossal magnetoresistance}
\author{Alan R. Bishop and Heinrich R\"oder\\
        Los Alamos National Laboratory\\
	MS B221, Los Alamos, NM 87545, USA, 
        \texttt{hro@karl.lanl.gov }}
\date{}
\maketitle
\abstract{
The history and recent developments in studying (colossal)
magnetoresistance in perovskite manganese oxides is reviewed.
We emphasize the growing evidence for strongly coupled spin,
charge and lattice degrees of freedom. Together with disorder,
these provide the microscopic driving forces for local and
inhomogeneous  textures. The modeling and experimental probes
for localized charge--spin--lattice (polaron) structures, and their
multiscale ordering, is discussed in terms of a growing synergy of
solid state physics and materials science perspectives.
}

\section*{Introduction}
It has been known since the early 1950's 
that manganese oxides,
when doped, become ferromagnetic metals, and exhibit 
remarkable magnetoresistive properties. Zener \cite{zen51} explained
magnetism in these materials via the \textbf{d}ouble
\textbf{e}xchange (DE) mechanism. He assumed that 
the only way charge transport
between $\Mn^{4+}$ and $\Mn^{3+}$ can happen is via the simultaneous
hopping of an $e_g$--electron from $\Mn^{3+}$ to the connecting $\O^{2-}$
and from the $\O^{2-}$ to the $\Mn^{4+}$ $e_g$ band;
hence the term double
exchange. Ferromagnetism is then induced via this hopping by
the very large Hund's rule coupling between the $\Mn$ $e_g$ and
$t_{2g}$ electrons, resulting from the high spin state of the manganese
$d$--electrons. An explicit formulation of this mechanism was first 
presented by Anderson and Hasegawa \cite{ah55}, and treated in a
mean--field type description by deGennes \cite{deg59}. 
In those works, an effective
one--band Hamiltonian is proposed which in the limit of
infinite Hund's rule coupling reads
\begin{equation}
H_{DE} = - \sum_{ij} \frac{t_{ij}}{\sqrt{2}} 
\sqrt{1+\frac{\mathbf{S}_i . \mathbf{S}_j}{S^2  }}
\left(
c_{i}^{\dagger}c_j^{\phantom{\dagger}} + H.c. 
\right) \;\;,
\end{equation}
describing the hopping of spinless fermions with
a hopping amplitude depending on the relative orientation of the
total $d$--electron spin on neigbouring sites $i$ and $j$. 
In early neutron scattering measurements \cite{wk55} and from
magnetic measurements \cite{jon56} the magnetic and structural phase
diagram of $\La_{1-x}\Ca_{x}\Mn\O_3$ was mapped out, with little
emphasis on the electric properties.

Kubo and Ohata \cite{ko71} derived a mean field theory for
the lanthanate manganites starting from the 
ferromagnetic Kondo lattice model
\begin{equation}
H_{KLM} = -t \sum_{ij,\sigma} 
\left(
c_{i,\sigma}^{\dagger}c_{j,\sigma}^{\phantom{\dagger}} + H.c.
\right)
-J_{H}\sum_{i,\sigma,\sigma'}
\left(
\mathbf{S}_i . \mathbf{\sigma}_{\sigma \sigma'}
\right)
c_{i,\sigma}^{\dagger}
c_{j,\sigma'}^{\phantom{\dagger}}
\;\;,
\end{equation}
where spin--$\frac{1}{2}$ $e_g$--electrons $c_{i,\sigma}^{\dagger}$ are
hopping in a cubic lattice coupled to a local $s=\frac{3}{2}$ spin, 
describing the three localized $t_{2g}$--electrons.
Kubo and Ohata calculate a magnetic phase diagram, the
resistivity and the magnetoresistance. Their results show a
ferro-- to para--magnetic phase transition at $T_c^m$, accompanied by
a change in the temperature dependence of the resistivity, and
a diverging negative magnetoresistance at this transition.
However, neither the predicted low--temperature dependence
of the resistance ($\approx T^{9/2}$) nor the constant
resistivity above $T_c^m$ agree with experiment. In a 
further paper \cite{ko72} they also qualitatively discuss the scattering of
the charge carriers from temperature  induced spin--disorder.

A recent shift of focus onto the magnetoresistive
behaviour of the manganates \cite{ksk89} has spawned
renewed interest in metallic peroskites $A_{1-x}' A_x \Mn\O_3$ 
($A' = \La,$$ \pr,$$ \Nd,$ ...,$ A =\Ca,$$\Sr,$$ \Pb $)
as a class, and  
a new and refined set of data from experiments using modern techniques
became available. The term \textbf{c}olossal
\textbf{m}agneto\-\textbf{r}esis\-tance (CMR) was coined to
distinguish these materials from the giant magnetoresistive
compounds, see e.g. \textit{apo96} and references therein. 
Typically it is found that at low (high) values
of doping $x$ the materials are antiferromagnetic insulators (AFMI)
at low temperatures and paramagnetic insulators (PMI) at high temperatures,
and at intermediate doping ($0.2 \lesssim x \lesssim 0.4$) a transition
from a low--temperature ferromagnetic metal (FMM) to a
PMI phase was found. It is in this region that the CMR effect occurs
most strongly. A sketch of the phasediagram is shown in Fig. 1.
Using a modern version of mean field
theory, the $d=\infty$ mean--field theory developed for
the treatment of correlated electrons in the high--$T_c$
superconductors, Furukawa \cite{fur95} reproduced the
earlier results of Kubo and Ohata.

In hindsight it appears surprising that little attention was
given to the possibility of local lattice or Jahn--Teller
distortions of the perovskite unit cells. The 
importance of a strong electron--coupling
in these and related perovskite was well--studied
\cite{kan60,goo61,goo64,kk73,kk82}, and found to be
important for both structural, magnetic and transport
properties in perovskites. The inclusion of a coupling of
the DE mechanism to the lattice degrees of freedom was
argued for by Millis \cite{mls95} and studied in detail
by us \cite{rzb96,zbr96} and Millis et al \cite{msm96,mms96:II}.

In this short review we will focus on the strong interaction
between spin, charge and lattice degrees of freedom with
special emphases on the possibility of inhomogeneous,
local effects, particularily charge localization, (small)
polaron formation and ordering. We believe those are the driving force
for many of the exciting phenomena exhibited by the
CMR perovskites, especially the formation of charge
and spin--ordered superlattices, the unusual magnetostrictive
behaviour, first order structural phase transitions driven by a
magnetic field, etc. 
In the next section  we gather the experimental evidence for polaronic
behaviour, mostly above $T_c^m$, followed by a  discussion of the
measurements of local structure and their relation to
macroscopically ordered phases. Then we describe the 
theoretical models and techniques used to include
lattice effects in an averaged way, and their possible
modifications and extensions to include local effects,
in spin, charge and lattice.
We conclude with a summary of some probable future
focal areas of research in the CMRs.
\section*{Polarons in CMR}
That polarons play an important role in the charge transport of the
CMR perovskites, at least in the insulating phase
above $T_c^m$ and the MI transition, is obvious from the
activated behaviour of the conductivity.
In addition, a coupling of the transport to lattice degrees of freedom 
is also evident from the
magnetostriction and magnetoelastic effects 
observed by Ibarra et al \cite{iam95}
in $\La_{0.6}\Y_{0.07}\Ca_{0.33}\Mn\O_3$.
Below a temperature $T_p = 320K$ they find a deviation of
the magnetostriction $\Delta l/l$ from normal behaviour down to
$T_c^m$ where $\Delta l/l$ drops to a lower value. 
In this regime the magnetostriction is isotropic. 
Ibarra et al interpret this as the region where ``large''
polarons start to form. The FMM phase is the low volume
phase and shows anisotropic
magnetostriction comparable to normal $d$--metals.
Even below $T_c$ there is evidence for at least
spin--polaronic behaviour. From a careful study
of resistivity $\rho$ versus magnetization $m$
in $(\La_{1-x}\Y_{x})_{2/3}\Ca_{1/3}\Mn\O_3$. Hundley et al \cite{hhh95}
Martinez et al \cite{mfs96} find the dependence
$\rho \propto \rho_m \exp(-m(h,t)/m_0)$. This
means that transport even in the FMM is controlled
by magnetic polarons. Martinez et al interpret the
localization of charge carriers above $T_c$ as being due
to scattering of these magnetic polarons on the 
magnetic disorder above $T_c$. However, it is certain that
the lattice is involved in the charge--localization, and
we believe that the lattice dynamics predominantly slaves the 
charge and spin dynamics and not vice versa.

The existence of small polarons above $T_c$ has 
been established by a variety of transport measurements.
For example the high temperature thermopower exhibits
typical polaronic behaviour \cite{jsr96,clj96,amt96,mtr96}.
The deduced activation energies fit well with the
ones obtained from Hall effect measurements \cite{jhs96}.
As $T$ approaches $T_c$ both the activation energies
for the resistivity and the thermopower become dependent
on a magnetic field showing the additional magnetic character
of these polarons. Jaime et al \cite{jsr96} therefore use the term
``magnetoelastic polarons'' to describe the carriers
just above $T_c$.

In addition to the above indirect evidence for the relevance
of polarons in the CMRs drawn from transport measurements 
there is now direct evidence.
Using electron paramagnetic resonance (EPR), which measures
the local magnetic response, 
Oseroff et al \cite{ots96} found activated behaviour in the
EPR resonance for $T>T_c^m$ in $\La_{1-x}\Ca_x\Mn\O_{3+\delta}$.
Already below about $T \lesssim 600K$ they find deviations
from standard Curie--Wei{\ss} behaviour. At temperatures
$20K$ above $T_c^m$ they deduce an effective spin 
of about 30. This means that about seven manganese
are involved in the magnetic character of the magnetoelastic
polaron. This agrees well with our unrestricted 
mean--field calculation
of the spin character of local polarons above $T_c$ \cite{rzb96}.
The additional influence of lattice degrees of freedom was
directly observed in the EPR measurements of
Shengelaya et al \cite{szk96} who studied
the influence of oxygen isotopic substitution
($^{16}\O / ^{18}\O$) on the EPR resonance. 
Shengelaya et al calculated  the strength of the
DE induced ferromagnetic interaction $J$
between $\Mn^{3+}$ and $\Mn^{4+}$ 
from the
temperature dependence of the EPR signal. They found
$J$ to  depend on the oxygen mass, which strongly indicates that
spin--lattice coupling is important , and indirectly means that
local charge transfer also depends on the lattice
since magnetism is induced by local hopping.

Direct measurements of the oxygen displacements are presented
by Sharma et al \cite{sxk96} from 
ion channeling (IC) experiments 
in 
$\Nd_{0.7}\Sr_{0.3}\Mn\O_3$,
$\pr_{0.7}\Ba_{0.3}\Mn\O_3$, and
$\La_{0.7}\Ba_{0.3}\Mn\O_3$. IC is
a direct real space probe of very small ($ \approx 1pm$) displacements $u$ of
atoms from their regular positions.
Sharma et al find a very strong correlation of the resistivity
with $\frac{d u}{dT}$, indicating that transport is strongly
correlated with \textbf{local} lattice distortions. As the
temperature is lowered from the PMI to the FMM phase 
the measured distortion decreases and reaches a steady
state below T_c. Both the $\Mn$ and to a lesser 
degree the rare earth sites are involved.

Summarizing these experimental results, we believe that
magnetoelastic polarons dominate transport above $T_c$.
As $T$ approaches $T_c$ from above, the magnetic component
of these composite
 particles becomes spatially extended, and eventually leads to
a delocalization of charge and spin degrees of freedom in the FMM phase.

\section*{Local structure measurements, polaron lattices  and charge ordering}
The experimental evidence for spatially inhomogeneous behavior in CMR materials
has begun to grow rapidly both in more conventional global signatures
(e. g. optics and thermodynamic properties, such as specific heat, by
magnetotransport measurements, Hall effect measurements, even within 
powder neutron measurements\cite{dzm96}) and. most strikingly in more local
probes. The latter include neutron 
pair--distribution function (PDF)analysis \cite{bdk96,lebr96}, 
extended x--ray fine structure (XAFS) \cite{tmc96},
 $\mu$SR \cite{hlh96}, XAS and ARPES \cite{pds96},
measurements of
the isotope effect \cite{zck96}, perturbed angular correlation spectroscopy
\cite{cea96}, and high--resolution electron
microscopy HREM \cite{htc96}. The availability of the local probes is in
many cases rather new and for the first time opens the possibility of
systematically exploring the interrelated reflections of multiscale
structure and dynamics in complex electronic materials such as transition 
metal perovskites. As we have emphasized, in those materials, the 
multiscale patterns may well have major intrinsic components and 
be driven at a microscopic level by strong coupling of spin, charge
and lattice degrees--of--freedom. The signatures of mesoscopic multiscale
patterns must then be measured in each of these degrees of freedom 
and shown to be self--consistent with each other to establish a valid 
microscopic model, and permit reliable predictions of macroscopic 
functionalities. While we are at the beginning of this systematic exploration
of transition metal perovskites (and other complex electronic materials), some
important recent results in CMR materials are already available. In
particular the resolution of fine--scale lattice structures and their
relation to charge (polaron)--localization and ordering represents a
quite new merging of traditional solid state and materials science
perspectives, and is gradually being augmented by 
electronic structure and magnetic signatures.

For example, Ramirez et al. \cite{rsc96} report thermodynamic
and electron diffraction signatures of charge-- and spin--ordering
in $\La_{1-x}\Ca_x \Mn\O_3$  around $x \approx 2/3$, 
where strong lattice commensurability
can stabilize a near--perfect superlattice ordering. They infer
(from specific heat and electron diffraction) charge--ordering at
$T_c \approx 260K$ accompanied by a large ($>10\%$) increase in the 
sound velocity, implying a significant electron--lattice coupling.
More direct elastic constant measurements are certainly needed for specific 
assignments of structural distortions. 

An alternate regime of (hole) charge--ordering occurs for 
$x \lesssim 0.5$ in $\pr_{1-x}\Ca_x \Mn\O_3$. This has been studied 
intensively by Tomioka et al \cite{tak96} taking additional
advantage of tuning via a magnetic field, and by Lees et al
\cite{lbb96} in less detail but over a wider concentration
regime. On the basis of resistivity measurements Tomioka et al 
suggest a real--space ordering at $T_c \approx 230K$ of 
$1:1$ $\Mn^{3+}:Mn^{4+}$ species optimized around the $x=\frac{1}{2}$ 
commensurate doping with discommensurations
growing in for $x<\frac{1}{2}$, weakening the insulating charge--ordered
state and eventually melting it ( in a strong enough magnetic field).
This melting is controlled by an external magnetic field and
a $x$--$H$--$T$ phase diagram is mapped out by Tomioka et al \cite{tak96}
Since the conduction electrons (polarons) are unusually strongly coupled to
the local spins ( via the DE effect), the resistive state is 
strongly affected by the local spin configuration. Similar
charge--orderings occurs in $\La_{1-x}\Ca_x \Mn\O_3$ \cite{xmg96,cc96},
$\pr_{1-x}\Sr_x \Mn\O_3$ \cite{tak96}, 
$(\Nd,\Sm)_{1/2}\Sr_{1/2} \Mn\O_3$ \cite{tkm96}, and 
$\Nd_{1-x}\Sr_x \Mn\O_3$ \cite{ktm95} but are
more confined to the near--commensurate $x=1/2$ doping. 
$\La_{1/2}\Ca_{1/2} \Mn\O_3$ is distinctive since it shows at least
four, with respect to spin-- and charge--distribution, 
phases as a function of magnetic field  \cite{xmg96}. It is
likely that the special role of $\pr_{1-x}\Ca_x \Mn\O_3$ is 
controled by the similarity in ion radius of $\Pr^{3+}$ and
$\Ca^{2+}$, which gives rise to less cation induced lattice
disorder \cite{rma96} and hence enables the observation of a  variety of
charge and spin ordered phases. The charge ordering in 
$\pr_{1-x}\Ca_x \Mn\O_3$ is accompanied \cite{tim96} by a change in the 
average lattice parameters of $0.6$ -- $2.4\%$ followed by
antiferromagnetic and canted antiferromagnetic ordering at
lower temperatures. Teresa et al \cite{tim96} have recently 
reported evidence for charge localization ( in the form of 
a small polaron) in $\pr_{2/3}\Ca_{1/3} \Mn\O_3$ 
for $T_p \lesssim 400K$ which,
however, only charge--orders below $T_c \lesssim 210K$. They report
thermal expansion
and magnetostriction measurements ( up to 14 T). A continuous
change in volume is measured for $T_c < T < T_p$, and the magnetic field
suppresses charge--ordering below $T_c$ giving rise to a first--order
structural transition. The electrical behaviour is similar to the
structural behaviour, again supporting a spin--charge--lattice
coupling.

 Evidence for coupled lattice and magnetic effects have been
reported by Argyriou et al \cite{amp96} in 
$\La_{0.875}\Ca_{0.125} \Mn\O_{3+\delta}$, following earlier reports of
a discontinuous volume contraction of the lattice ($\Delta V/V \approx 0.1\%$)
at the ferromagnetic transition temperature.
Based on powder neutron and
Rietveld analysis, Argyriou et al suggest that as $T$ is lowered below
room temperature there is a a rapid development of a large breathing--mode
(oxygen) distortion coinciding with large positive and 
negative expansions of the $c$ and $b$ axes, respectively. 
Canted ferromagnetic ordering onsets at $\approx 220K$ 
correlated with the reduction of the (averaged) breathing distortion.
Similar effects were reported by Kawano et al \cite{kkk96} in
$\La_{1-x}\Sr_{x} \Mn\O_{3}$ ($x<.17$), who discuss these effects
in crystallographic terms. However, the maximum shift of the
scattering peaks at $x=\frac{1}{8}$ makes a discussion in terms of
an ordered polaron superlattice more appropriate. Indeed in the same
material Yamada et al \cite{yhn96} find a polaron lattice
at $x=\frac{1}{8}$ accompanied by high resistivity. At nearby
concentrations one might expect the system to phase separate
into the ordered $x=\frac{1}{8}$ phase, or, more likely, 
defects will apear in the ordered phase as in other
commensurate to incommensurate transitions.

Turning from these more global signatures of charge localization and
ordering, and their coupling to spin and lattice, it is especially
exciting that \textbf{local} probes are now beginning to yield direct
information on polaron formation and ordering. Early XAFS
measurements \cite{tmc96} clearly demonstrated 
that the $\Mn\O$ distances developed multiple length-scales above
$T_{MI}$ consistent with small polaron formation in a homogeneous
background matrix. Careful neutron PDF measurements \cite{bdk96} in
$\La_{1-x}\Ca_{x} \Mn\O_{3}$ ($x=0.12,0.21,0.25$) 
have supported and significantly 
expanded this result. Billinge et al \cite{bdk96} found 
a large change in the
local structure connected with the metal--insulator 
transition for $x=0.21,0.25$ 
(and no such change in the $x=0.12$ sample which exhibits no metal--insulator
transition). The local structure change is plausibly modeled as an
isotropic collapse of oxygens toward the $\Mn$ of magnitude 
$\delta=0.12 A$ occuring on one in four $\Mn$  sites. Importantly, these
authors also find a large ($\Delta T \approx 50K$) temperature range
\textbf{below} $T_{MI}$ where strongly 
non--thermal $\Mn\O$ distortions persist,
indicative of precursor fine--scale lattice distortions. Similar
anomalous deviations from harmonic  Debye--Waller behaviour were also
observed in $\La_{0.65}\Ca_{0.35} \Mn\O_{3}$ by
a careful reanalysis (following the increasing evidence for
unusual lattice effects) of powder neutron studies \cite{dzm96}. The 
evidence for local structural defects is further illuminated 
by the neutron PDF work of Louca et al \cite{lebr96} on
$\La_{1-x}\Sr_{x} \Mn\O_{3}$ ($x = 0 - 0.4$). 
These authors demonstrate that 
the local atomic structure deviates from the average
crystallographic structure throughout this $x$--range, including
both the orthorhombic and rhombohedral crystal structures;
the crystallographic rhombohedral structure shows no
signatures of distortions of $\Mn\O_6$ octahedra, see Fig. 2. The PDF signatures
for local Jahn--Teller distortions occur as small (hole) polarons in the
in the paramagnetic insulating phase and persist as local ( but more
extended) polaron distortions below the transition in the ferromagnetic
metallic phase, resulting in percolative and inhomogeneous metal
signatures for transport and magnetic properties.

The notion of \textbf{local} JT distortions even in the rhombohedral
and metallic phases is entirely consistent with the angle--resolved
photoemission and XAS park et al reports of \cite{pds96}, and 
with the large isotope ($^{16}\O/^{18}\O$) dependence of
$T_c$ \cite{zck96}.

Finally we note the exciting beginning of direct HREM imaging of
lattice distortions in microdomains of charge localized carriers.
Hervieu et al \cite{htc96} have studied $\pr_{0.7}\Ca_{0.25}\Sr_{0.05} \Mn\O_{3}$ 
and $\pr_{0.75}\Sr_{0.25} \Mn\O_{3}$ and been able to directly image monoclinic
domains in a $\Gd\Fe\O_3$--type background matrix . They suggest 
 strong local modifications of the $\Mn\O_6$ octahedral structure around
localized charges ($\Mn^{4+}$). These localized lattice and charge distortions
are observed to cluster into microdomains of short
linear segments which may be viewed as $(A^{2+}\Mn^{4+}\O_3)_n$ embedded in 
the mixed valent $\Mn^{3+}$--$\Mn^{4+}$ perovskite matrix. Presumably these
microdomains will order into charge--ordered superlattice phases 
near commensurate dopings.
\section*{ Theoretical approaches:}
As pointed out above initially theoretical 
investigations focused on the study of the 
DE mechanism. The early mean--field theories
\cite{deg59,ko71,ko72} and even the 
$d=\infty$ mean--field theory of Furukawa \cite{fur95}
all describe the FMM as a simple ferromagnet, and the
conduction as that of a simple metal. However, as early as 
1982 Kubo \cite{ku82} pointed out that the problem is
not that simple. From exact diagonalization in  a one dimensional 
model he noticed
that even the ground state of the
ferromagnetic Kondo lattice (FKLM) may be unusual 
(for periodic boundary conditions). He also points out that
even if there is a simple ferromagnetic ground state,
there are other states closeby  in energy, which would
give rise to unconventional excitations. This work
was extended by Zang et al \cite{zrb96}, in which 
small 2 and 3 dimensional systems , and their
excitations were studied in detail. It was found that
only for very special concentrations is a simple ferromagnet indeed
the groundstate of this model. Th excitation spectrum
of the KLM model shows atypical spin--wave disperions for
large wave vectors. This new and unusual excitations 
are born out experimentally in the unusual spin dynamics
seen the $\mu$SR experiments by Heffner et al \cite{hlh96},
and by the surprising incoherent spin wave scattering
observed by Lynn et al \cite{leb96}. 
The $\mu$SR experiments \cite{hlh96} show clearly that
there are slow spatially inhomogeneous spin dynamics
below $T_c$.
However, there is
substantial experimental controversy; e.g. Perring et al. \cite{pa96}
find perfect agreement of the spin wave dispersion of
a simple nearest neighbour ferromagnetic Heisenberg model
with their neutron scattering data. In light of the 
importance of local effects on structure and transport, it
would be extremely desirable to have similar data for
spin structures, for example via spin--polarized neutron
PDF. Being faced with the problem that all approximate treatments
of the FKLM are inconclusive R\"oder et al \cite{rsz96} 
performed a high temperature series analysis of the FKLM.
This analysis clearly shows a transition at the
experimentally observed values. However, the concentration
dependence of this transition resembles the one obtained
for a ferromagnetic ferrimagnet. This might explain the
perfect spin--wave disperion observed by  Perring et al. \cite{pa96},
since the lower branch of a ferrimagnets dispersion is
exactly like the one of a nearest neighbour Heisenberg model.
In addition, a ferrimagnetic ordering would imply a
tendency of the purely electronic FKLM towards
charge ordering and charge localization. This pre--formation
of charge-ordering renders the FKLM very susceptible
to the formation of ordered polaron states, when
lattice effects are included. 

Notwithstanding the theoretical difficulties arising from
the complicated many--body effects of the FKLM
Kondo lattice model, a coupling of charge and lattice
degrees of freedom is necessary both on theoretical and
experimental grounds. In a theoretical treatment one
necessarily has to resort to various mean--field theories.

In a dynamic mean--field theory Millis et al \cite{msm96,mms96:II}
treat the DE coupling dynamically, but use the
frozen--phonon approach for the lattice. Due to
pecularities of the dynamic mean--field theory ($d=\infty$) they
only treat extreme densities $x=0$ and $x=1$. As a result
of this calculation they observe a metal--insulator
transition driven by the localization of electrons on
disorder induced by lattice--fluctuations. They
mimic the $x$--dependence by a variation in the 
strength of the Jahn--Teller (electron--phonon) coupling
constant. Unfortunately this description does not
allow the inclusion of local structure effects due to
the limitations of the dynamic mean--field theory.

Th approach taken by R\"oder et al \cite{rzb96,zbr96} is in some
sense complementary, since it uses standard mean--field
theory for the DE coupling. This assumes that the electrons
move in a static field created by the spin (this 
approximation is correct in the classical limit), but
allowing for phonon dynamics by using a well--established
variational scheme. Their results agree qualitatively
with the ones by Millis et al., but allow a more systematic
study of the concentration dependence. It should be possible
to actually use the treatment of the phonons from \cite{rzb96,zbr96}
in the ($d=\infty$) theory, and also to extent the
calculation of \cite{msm96,mms96:II} for more realistic
fillings. Such a theory would show the dependenc of
the tranition on the various time scales arising
from charge, spin and lattice degrees of freedom.

Unfortunately, if one is interested in local structure
effects, it is necessary to resort to the approximations
in  \cite{rzb96,zbr96}. As pointed out before such an
unrestricted -- in the sense that we allow the
field variables to be spatially inhomogeneous --
calculation explicitly shows the temperature dependence
of the spatial of the various fields involved.
This question is especially interesting closely
above $T_c$. The coupling of the lattice polarons
to the magnetic degrees of freedom via the
DE interaction leads to the phenomena of
entropic localization. Magnetoelastic
polarons consist  of a fairly localized
lattice distortion, a localized charge and a 
ferromagnetic bubble which extends from the
center of the polaron over the neigbouring sites.
As the temperature increases the entropy in 
the disordered spin background becomes
more important and compresses the ferromagnetic
bubble, leading to an increasing localization
with temperature. This picture is a 
single particle description and needs to be modified
if the density of polarons becomes large. In
perliminary studies we have investigated the
interaction of two such magnetoelastic polarons ,see Figs. 2,3.
As the temperature is lowered  at high 
temperature localized polarons interact via
their magnetic part, wich is of longer range than
either their lattice or charge constituents,
and form a variety of bound states with ever--increasing
equilibrium radius. Within these large bound multipolarons  first
spin and then charge decouple from the lattice
degrees of freedom, which might eventually lead to the
breakdown of the magnetoelastic polaron below
the transition.  These studies need to be extended in
various directions.

For the explanation of the measured isotope effect 
\cite{zck96} it is necessary to remain in a two--orbital
description. If one wants to project onto a single
orbital description it is necessary to resort to
phenomenological changes in the electron phonon
coupling constant with isotopic substitution. In the
two--orbital description this effect is more transparent. 
   
The nature of the lattice distortion arises from
two sources. As pointed out by Millis \cite{mil96} in
addition to Jahn--Teller distortions caused
by the splitting  of the occupied $e_g$--levels in a
cubic field, induced holes would lead to breathing
like distortions which would frustrate the Jahn--Teller
ordering. This may give rise to an order--disorder
transition involving the Jahn--Teller distorted 
octahedra. It is important to include the relative dynamics
betweeen the various lattice modes; the breathing modes may not
to a localization at all, depending on how slow or
fast the respective lattice relaxations are.
This physics may be included using the
microscopic interaction Hamiltonian in \cite{mil96}
and the unrestricted mean field approach from \cite{rsz96}.
However, even more important may be static disorder
induced by the distribution of cations of different 
radii, as first pointed out in \cite{rma96}. If there
are both frustration and disorder present one might
expect glasssy behaviour with respect to local
modes, and not caused by frustrated spin interactions,
as in spin glasses. Such lattice glas models 
may hinder the freezing of the lattice in
commensurate and therefore charge ordered states
and lead eventually to the charge delocalization 
as in the FMM phase.

It is important to note that the scattering of
the electrons on spin disorder in the paramagnetic
phase does not lead to a localization of the
electron wave function at realistic concentrations
\cite{lzb96}. Therefore Anderson localization
cannot explain the observed activated behavior 
above $T_c$.

\section*{Future directions}

As in other transition metal perovskites, the CMR manganite oxides represent
a class of complex electronic materials where strong coupling of spin, charge
and lattice degrees--of--freedom is the essential ingredient. These
microscopic couplings, together with disorder (e.g. from dopant ions), result
in competing interactions (disorder plus frustration) which drive 
multiscale textures and associated (glassy) dynamics. The macroscopic 
functionalities (CMR, ferroelectric, superconducting, etc.) are 
determined by this mesoscale complexity. Thus understanding the 
microscopic--mesoscopic--macroscopic relationships will be a
prevailing research theme in these and other complex electronic materials, as 
a route to providing scientific principles for controlling 
synthesis--structure--property relationships.

We have emphasized that if this strategy is to yield a 
validated and predictive modeling capability, it will require
\textbf{correlated} probing of spin, charge and lattice properties 
on multiple length and timescales, including  local scales. The complexity
manifests itself as fine--scale structure (twinning, tweed, 
microdomains) and elasticity in materials science, but
as charge localization and ordering (polarization, polaron
formation, and superlattices) in solid state physics. Correspondingly, 
modeling must focus on minimal models which respect spin, charge and lattice,
and incorporate distinctly nonlinear, nonadiabatic, and discrete lattice
effects. As we have emphasized, both experiment and theory are now
faciliatating this agenda, and this will surely mark future 
research on CMR in manganite oxides and related
materials. The materials, their intriguing functionality (CMR),
and new experimental and theoretical techniques have converged
fortuitously to make this an exciting new era for studying
small polaron physics -- in the generalized coupled spin,
charge, lattice sense we have discussed here -- and the intrinsic
fine scale structure associated with solid--solid (elastic) 
phase transformations. This synergy of solid state physics and
materials science perspectives holds rich promise.

The issue of time scales is worth emphasizing. This has long
been a fundamental concern for fine--scale 
precursor structure
at solid--state phase transformations, especially of elastic, 
martensitic character. Is the
JT distortion static or dynamic or both?  Are their cluster dynamic time scales
giving rise to  central peaks in the dynamic structure factor? What is
the dynamics of nucleation? Controversies surrounding these
questions have a confusing history (see e.g. \cite{mg91,sys89} and
references therein), because
time scale probes have been very limited. The expectation of multiple
spatial scales
for the mesoscopic octahedral distortions in fact leads to expectations
of multiple temporal scales, of even hierarchical or glassy character
There are some preliminary
suggestions of such glassy responses in 
recent $\mu$SR data \cite{hlh96}, as well as controversial
neutron dynamics results (see above) \cite{dzm96,bdk96}. This certainly
will be another focal area for future experimental research.

\textbf{Acknowledgements:} We would like to thank Stuart Trugman, Bob Heffner,
Myron Salamon, Ekhard Salje, Takeshi Egami, Despina Louca, Andy Millis,
Rajiv Singh,and Jun Zang for many helpful and illuminating discussions.

\bibliography{cmr}
\bibliographystyle{ieeetr}

\newpage
\begin{figure}
\caption{A sketch of the phasediagram of the CMR materials. The lines
are only guides to the eye. The various phases are:
paramgnetic insulators (PMI), antiferromagnetic insulators (AFI),
ferromagnetic metal (FMM), antiferromagnets (AF), and charge--ordered (CO). }
\end{figure}
\begin{figure}
\caption{The Mn--O bondlengths as determined from the the PDF (triangles)
compared to those deduced from the lattice constants of the 
crystal structure (solid lines L, S1 and S2). Note that the
local distortion is present even in the rhombohedral phase. From
\cite{lebr96}. }
\end{figure}
\begin{figure}
\caption{ The charge (a.)  and spin (f.)  configuration of 
two interacting magnetoelastic polarons. The picture is
a cross section from a $30^3$ system with periodic boundaries.
The greyscale indicates the charge and spin distribution,
respectively. Note that the charge is more localized
than the spin. }
\end{figure}
\end{document}